\documentclass[10pt,twoside]{article}
\usepackage{amsfonts}
\usepackage{fancyhdr}
\usepackage{titlesec}
\usepackage{cite}
\usepackage{ifthen}
\usepackage{amssymb}
\usepackage{pifont}
\usepackage{stmaryrd}
\usepackage{setspace}
\usepackage{indentfirst}
\usepackage{amsmath,amssymb,amscd,bbm,amsthm,mathrsfs,dsfont}
\usepackage{graphicx}	
\usepackage{stackrel}
\usepackage{algorithmic}
\usepackage{float}
\usepackage{subcaption}
\usepackage[colorlinks=true, allcolors=blue]{hyperref}	
\usepackage{verbatim}
\usepackage[table,xcdraw]{xcolor}
\usepackage{array}
\usepackage{dcolumn}
\usepackage{algorithm} 
\input amssym.def

\titleformat{\section}{\centering\large\bfseries}{\S\arabic{section}}{1em}{}
\newboolean{first}
\setboolean{first}{true}

\usepackage[normalem]{ulem}
\useunder{\uline}{\ul}{}

\def\BibTeX{{\rm B\kern-.05em{\sc i\kern-.025em b}\kern-.08em
    T\kern-.1667em\lower.7ex\hbox{E}\kern.125emX}}
 
\begin{document}

\setlength\abovedisplayskip{2pt}
\setlength\abovedisplayshortskip{0pt}
\setlength\belowdisplayskip{2pt}
\setlength\belowdisplayshortskip{0pt}

\title{\bf \Large A Region of Interest Focused Triple UNet Architecture for Skin Lesion Segmentation

\author{Guoqing Liu$^{1}$, Yu Guo$^{1}$,  Caiying Wu$^{1}$, Guoqing Chen$^{1}$,  \\  Barintag Saheya$^{2}$, Qiyu~Jin$^{*,1}$
\date{2023}}}
\maketitle

\footnote{Supported by National Natural Science Foundation of China (No. 12061052,  62161044), Young Talents of Science and Technology in Universities of Inner Mongolia Autonomous Region (No. NJYT22090), Natural Science Fund of Inner Mongolia Autonomous Region (No. 2020MS01002), Innovative Research Team in Universities of Inner Mongolia Autonomous Region (No. NMGIRT2207),  Prof. Guoqing Chen's “111 project” of higher education talent training in Inner Mongolia Autonomous Region and the network information center of Inner Mongolia University.}
\footnote{* Corresponding author}

\begin{center}
\begin{minipage}{135mm}
{\bf \small Abstract}.\hskip 2mm {\small 
Skin lesion segmentation is of great significance for skin lesion analysis and subsequent treatment. It is still a challenging task due to the irregular and fuzzy lesion borders, and diversity of skin lesions. In this paper, we propose Triple-UNet to automatically segment skin lesions. It is an organic combination of three UNet architectures with suitable modules. In order to concatenate the first and second sub-networks more effectively, we design a region of interest enhancement module (ROIE). The ROIE enhances the target object region of the image by using the predicted score map of the first UNet. The features learned by the first UNet and the enhanced image help the second UNet obtain a better score map. Finally, the results are fine-tuned by the third UNet. We evaluate our algorithm on a publicly available dataset of skin lesion segmentation. Experiments show that Triple-UNet outperforms state-of-the-arts on skin lesion segmentation.}
	\end{minipage}
\end{center}

	\section{Introduction}
		The American Cancer Society predicts that there will be $108,480$ new cancer cases and $11,990$ cancer deaths to occur in the United States \cite{2022Cancer}, therefore how to diagnose and treat skin cancer quickly becomes a very important task. Skin lesion segmentation is of great significance for improving the quantitative analysis of skin cancer, and hence it is the critical step in skin cancer diagnosis and treatment planning. Manual delineation is, however, usually tedious, time-consuming, and error-prone. In clinical applications, the high requirements of automatic segmentation technology are adopted to improve the efficiency and accuracy of skin lesion segmentation. This, however, inevitably leads to sharp attention and gives rise to an enormous challenge. It is mainly because of the following aspects: (1) Skin lesions vary greatly in size, shape, and color, and some are even obscured by hair. (2) The boundaries between some lesions and normal skin are ambiguous. (3) The limited high-quality labeled images make the training even harder \cite{2020Polyp}. These all cause trouble for skin lesion area automated segmentation.
		
		Albeit traditional algorithms \cite{2017arXiv170207333A,2017arXiv170303186R,2017arXiv170304301J,2018arXiv180806759P,2018arXiv181000871I,2020arXiv201004022K} based on various hand-crafted features are interpretable, they have relatively poor stability and robustness owing to that each hand-crafted feature cannot perfectly characterize distinctive representations of skin lesions. Then, it leads to relatively poor segmentation performance for lesions with large variations. To solve this issue, data-driven algorithms based on convolutional neural networks (CNNs), such as Full Convolution Neural Network (FCN) \cite{2015FCN} and UNet \cite{2015UNet}, have been proposed to learn distinctive representations of skin lesions and their performance has been greatly improved compared to the traditional ones \cite{2010Auto-Context,2017AutomaticSL}.
		
		However, these data-driven models are still insufficient to tackle skin lesion segmentation due to large variations of skin lesions, ambiguous boundaries, and the limitation of high-quality images and labels. With regard to this, there are two main ways to solve this problem: enhancing image data and designing a more robust and efficient CNNs model. Simple image enhancement technologies have little help to the task of image segmentation because most of them focus only on local pixel-level information such as color and texture and do not involve global image-level information. The combination of local and global clues makes skin lesion segmentation more reliable. 
		The CNNs model extracts not only the local pixel-level information of the image but also the global image-level information of the image. From this point of view, it can get richer features. Moreover, in terms of segmentation tasks, we only need to strengthen the skin lesion area. Therefore, we use the predicted score map of the CNNs model to highlight the skin lesion area and make it easier to segment. Then, we improve the segmentation performance of the network.
		
		The main contributions of this work are:
		\begin{itemize}
			\item A novel architecture is proposed for skin lesion segmentation, which contains three U-shaped architectures. All sub-networks in the network are built from scratch and have not been pre-trained.
			\item A region of interest enhancement module (ROIE) is proposed to enhance the region of interest (ROI) in the input image. By improving the quality of network input, the segmentation performance is significantly improved.
			\item The proposed Triple-UNet is evaluated on the ISIC 2018 skin lesion boundary dataset, and state-of-the-art performance was achieved.
		\end{itemize}
		
		The rest of the paper is organized into five sections. Section $2$ provides an overview of the related work. In section $3$, we describe the proposed architecture. Experiments and results are in Section $4$. Finally, we summarize the paper and discuss future work and limitations.
		
\section{Related Work}
\subsection{CNN Models}
Since FCN was proposed, CNNs have been the most advanced algorithm in image segmentation applications. However, different from ordinary image segmentation, medical images usually contain noise and fuzzy boundaries. Therefore, it is difficult to detect and recognize the targets in medical images only by the local pixel-level information of the image. At the same time, the accurate boundary can not be obtained only by the global image-level information. Ronneberger et al. \cite{2015UNet} proposed UNet, a convolutional neural network with an encoder-decoder structure. In UNet, the down-sampling operation in the encoder is utilized to reduce the resolution of the feature maps. By using the down-sampling operation, the network obtains different receptive fields and captures both local pixel-level and global image-level information. The skip connections effectively integrate different scales and levels of feature maps from the encoder and decoder. At present, UNet has become one of the neural network architectures widely used in medical image segmentation. Therefore, many variants \cite{2018UNet++,2018AttentionUNet,2019ResNet++,2020UNet+++,2020MultiResUNet,2020DCUNet} have been proposed for segmentation.
    
    \subsection{Application of Attention Mechanism}
        Attention mechanism has been regarded as an advanced technique to capture long-range feature interactions and to boost the representation capability for convolutional neural networks \cite{2021arXiv210808205B}. Non-local mean captures remote dependencies effectively by calculating the response at one location as the weighted sum of features at all locations. Hence, Wang et al. \cite{2018NLNet} introduced a non-local block to capture remote dependencies by computing an attention map by measuring the relationship of each pair of pixels. Later, Fu et al. \cite{2019DANet} proposed a new attention module with two dimensions of the channel and space based on the non-local operation. It significantly improved the segmentation results. Unfortunately, it consumes a lot of resources due to the complexity of the non-local operation. To solve this problem, Huang et al. \cite{2020CCNet} gave a criss-cross attention module. For each position, the criss-cross attention module aggregates contextual information in its horizontal and vertical directions first, and then it applies another criss-cross attention module, finally, it enables the full image dependencies for all positions. However, there are still problems with heavy computation. Hu et al. \cite{2020SENet} proposed another form of attention, the Channel Attention Module, which adaptively recalcitrates the channel-wise feature responses by explicitly modeling interdependencies between channels. It greatly reduces the consumption of resources. Subsequently, Woo et al. \cite{2018CBAM} extended the design in \cite{2020SENet} to the spatial dimension, where the module generates attention maps along both the channel and spatial dimensions. However, it is very difficult for the generated spatial attention map to ensure that the region it focuses on is the target region of the task. Therefore, it is necessary to use a network used for medical image segmentation to learn where the target region of the task is.
    
    \subsection{Skin Lesion Segmentation}
		To apply deep learning techniques to skin lesion segmentation, a simple solution is to fine-tune and optimize the powerful CNNs appropriately \cite{2018ANF,2019BCDUNet,2020MCGUNet,2022MSRFNet}. Jha et al. \cite{2020DoubleUNet} proposed a new structure: DoubleUNet, which successively combined two U-shaped network architectures in a series. The result of the first U-shaped network is utilized to filter out the useless background in the image. The second U-shaped network is used to refine the segmentation results obtained by the first U-shaped network. However, due to the variety of skin lesions and fuzzy boundaries, the filtered pixels may not all belong to the useless background. Therefore, we incorporate the idea of residual learning into our design. Because the information of the original input is preserved, our proposed model achieves higher accuracy and generates better segmentation masks for challenging images. We evaluate the proposed Triple-UNet on the ISIC-2018 Lesion Boundary Segmentation \cite{2019Skin,2018The}, which is the largest publicly available dermoscopy image. Experimental results demonstrate that our architecture has significant performance improvement.
	
\section{Method}
\begin{figure*}[t]
		\centering
		\includegraphics[scale=0.5]{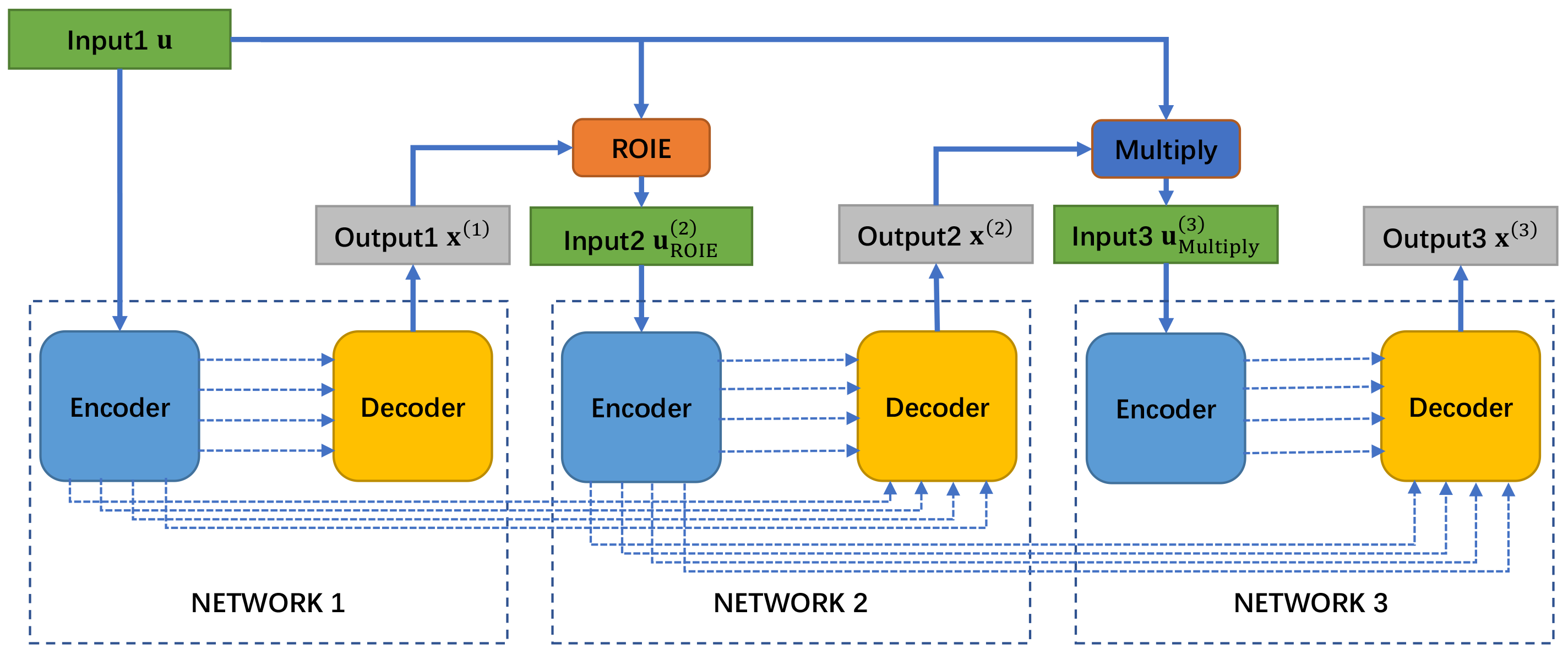}
		\caption{Block diagram of the proposed Triple-UNet architecture.}
		\label{Img1}
	\end{figure*}

    Let $\mathbf{u}$ be an image which is defined on a domain ${\rm \Omega} \in \mathbb{R}^{2}$ and $N=\mathrm{Card}({\rm \Omega})$ be the total number of pixels in the image $\mathbf{u}$. The segmentation task is to classify $\Omega$ into $2$ partitions: region of interest (ROI) partition $\mathbf{A} \subset {\rm \Omega}$ and background partition ${\rm \Omega}\setminus \mathbf{A}$. $\mathbf{x}$ is predicted score maps and its element $x_{i}$ represents the probability that pixel $i$ belongs to $\mathbf{A}$. Figure \ref{Img1} shows an overview of the architecture of the proposed algorithm. The structure is composed of three sub-networks with encoder and decoder structures. They are connected through different modules.

	In the first sub-network (NETWORK 1), the input image $\mathbf{u}$ is fed into a variant based on the UNet architecture to generate the predicted score map Output1 $\mathbf{x}^{(1)}$. Then, $\mathbf{u}$ and $\mathbf{x}^{(1)}$ are fed to the ROIE module (it will be discussed in detail in section \ref{Sec: Connectivity}). The ROIE module enhances the ROI partition in $\mathbf{u}$ through $\mathbf{x}^{(1)}$ to provide Input2 $\mathbf{u}^{(2)}_{\mathrm{ROIE}}$ as the input of the second sub-network (NETWORK 2). Finally, the predicted score map Output2 $\mathbf{x}^{(2)}$ generated via NETWORK 2 and is multiplied by $\mathbf{u}$ to get Input3 $\mathbf{u}^{(3)}_{\mathrm{Multiply}}$.  We use $\mathbf{u}^{(3)}_{\mathrm{Multiply}}$ as the input of the third sub-network (NETWORK 3) which generates a finer score map Output3 $\mathbf{x}^{(3)}$. To connect the three sub-networks more efficiently, the feature maps generated by the encoders in the previous sub-network are also fed into the decoders of the later sub-network as shown in the figure. The ground truth and the predicted score map are used to calculate the loss of each sub-network for providing deep supervision in each stage. The final image segmentation prediction map is the binarization of the third score map $\mathbf{x}^{(3)}$.
	
	In the following sections, we will first discuss the encoder and decoder in our network. It uses depthwise separable convolution and the channel attention module to improve the encoder and decoder in UNet. Then, Section 3.3 will introduce the connectivity structure among sub-networks in our network in detail.
	
    \subsection{Encoder}
	    Firstly, two depthwise separable convolution blocks, which are combined by a $3\times3$ depthwise separable convolution, batch normalization, and a Rectified Linear Unit (ReLU) activation function, in turn, are concatenated by the encoder per layer in each sub-network. They effectively improve the computational efficiency and reduce the model parameters, then allow the model to work on smaller devices and speed up image segmentation. Secondly, a channel attention module is placed behind these two convolution blocks. It redistributes the weights of the feature maps to improve the quality of feature maps generated by depthwise separable convolution blocks. Finally, max-pooling is performed with a pooling size of $2\times2$ and stride of $2$ to reduce the spatial dimension of the feature maps. Considering model specification and efficiency, the encoder uses $32$, $64$, $128$, $256$, and $512$ filters per layer, respectively. Note that these numbers are less than the number of filters for UNet and its variants, resulting in fewer parameters and computations.
	
    \subsection{Decoder}
        From Figure \ref{Img1}, our entire network has three decoders with slight modifications. Each layer of the decoder performs $2\times2$ bi-linear up-sampling on the previous decoder layer's feature map and receives the corresponding encoder layer's feature map by skip connections. As shown in Figure \ref{Img1}, the first decoder only has a skip connection with the first encoder, while the second and third ones use skip connection with the corresponding and previous encoders to ameliorate the accuracy of the output feature maps. After concatenation, a depthwise separable convolution of $3\times3$ is used to combine the low-level detailed feature maps from the encoder and the high-level semantic feature maps from the decoder. After the convolution layer, we use the batch normalization layer and the ReLU activation function. After that, we use the channel attention module to improve the quality of feature maps. At the end of the decoder, the score map is generated via a convolution layer with a sigmoid activation function.
		
    \subsection{Connectivity structure among sub-networks}
    \label{Sec: Connectivity}
\begin{figure*}[t]
    			\centering
    			\includegraphics[scale=0.5]{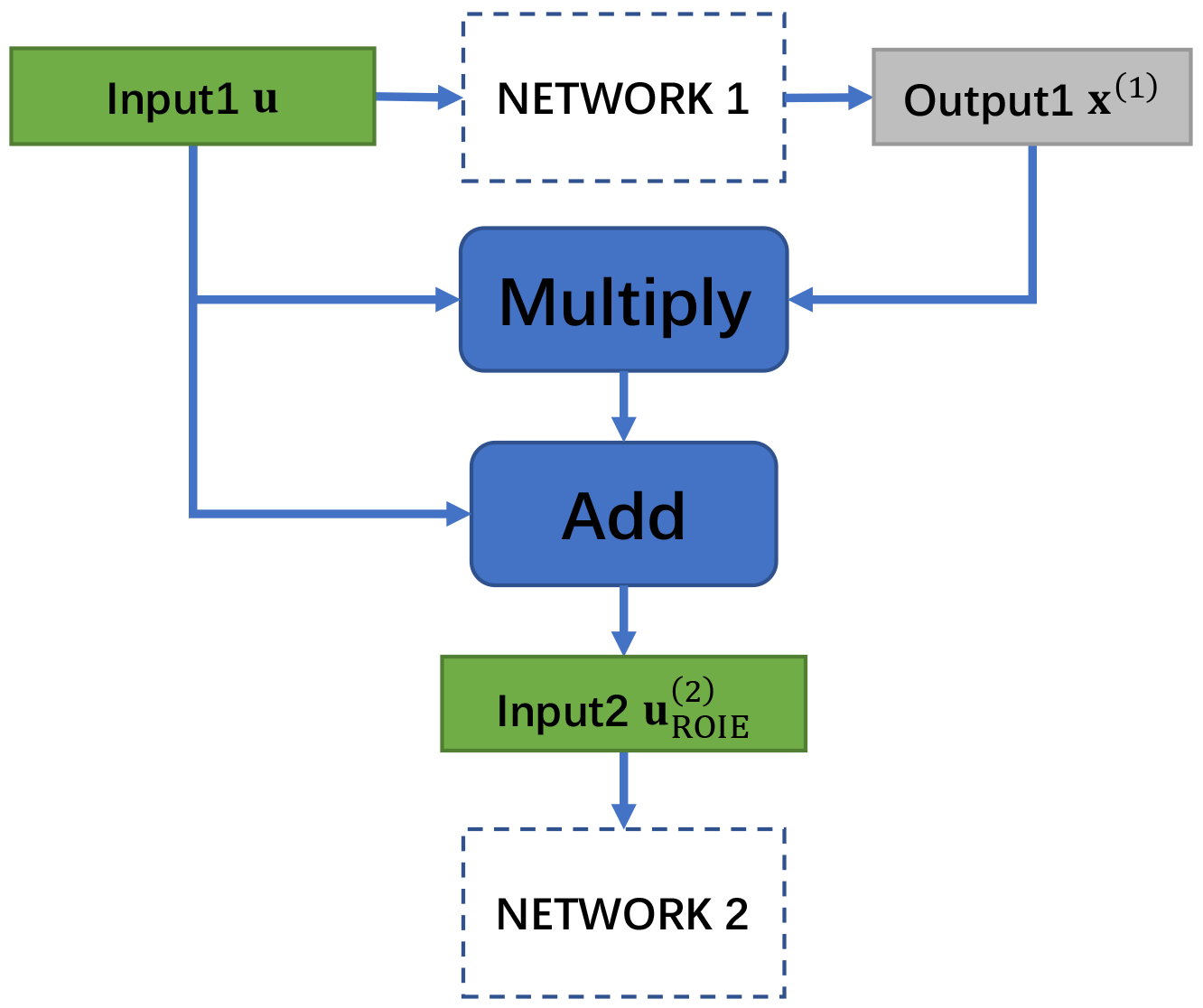}
    			\caption{The architecture of ROIE block.}
    			\label{Img2}
    	\end{figure*}

        As shown in Figure \ref{Img2}, the ROIE module first performs point-wise multiplication between Output1 $\mathbf{x}^{(1)}$ and Input1 $\mathbf{u}$, and then add the result to $\mathbf{u}$ to obtain Input2 $\mathbf{u}^{(2)}_{\mathrm{ROIE}}$, i.e.
        \begin{equation}
        \label{eq: ROIE}
            \mathbf{u}^{(2)}_{\mathrm{ROIE}} = \alpha \mathbf{x}^{(1)} \odot  \mathbf{u} + \beta \mathbf{u} ,
        \end{equation}
        where $\mathbf{x}^{(1)} \odot \mathbf{u}$ denotes the point-wise product between $\mathbf{x}^{(1)}$ and $\mathbf{u}$, $\alpha>0$ and $\beta>0$ are super parameters, which are set to $1$ according to the experiment. The term $\mathbf{x}^{(1)} \odot \mathbf{u}$ preserves the pixels belonging to the target object and suppresses the pixels belonging to the background. Since NETWORK 1 cannot achieve completely accurate segmentation and some pixels belonging to the target object in $\mathbf{u}$ are not accurately classified, these pixels are weakened after multiplying $\mathbf{u}$ by $\mathbf{x}^{(1)}$ compared with $\mathbf{u}$. Therefore, we add the result of $\mathbf{x}^{(1)} \odot \mathbf{u}$ to $\mathbf{u}$ to get $\mathbf{u}^{(2)}_{\mathrm{ROIE}}$ for further segmentation. The new input $\mathbf{u}^{(2)}_{\mathrm{ROIE}} $ is an enhanced version of $\mathbf{u}$, which contains both the enhanced target object area, the unenhanced target object area, and the background area. Then, NETWORK 2 is used to carry out a second segmentation on $\mathbf{u}^{(2)}_{\mathrm{ROIE}} $ and improve the output feature maps generated by NETWORK 1. If the problematic pixels are classified into the wrong category by NETWORK 1, NETWORK 2 reclassifies them to get a better-predicted score map.

    	\begin{figure*}[t]
    	    \begin{center}
    	    \addtolength{\tabcolsep}{0pt} 
    	    {%
    		\fontsize{8pt}{\baselineskip}\selectfont
    		\begin{tabular}{ccc}
    		\includegraphics[width=0.22\textwidth]{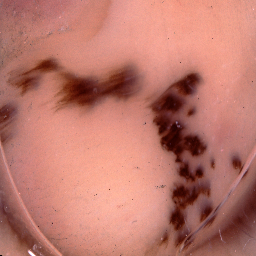} &
    		\includegraphics[width=0.22\textwidth]{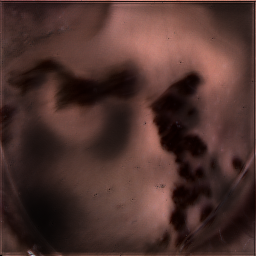} & 
    		\includegraphics[width=0.22\textwidth]{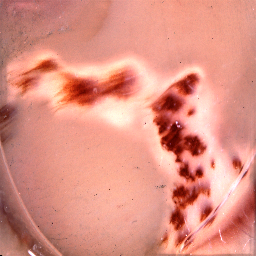} \\
    		(a) Input1 & (b) Input2(DoubleUNet) & (c) Input2(Ours) \\
    		
    		\includegraphics[width=0.22\textwidth]{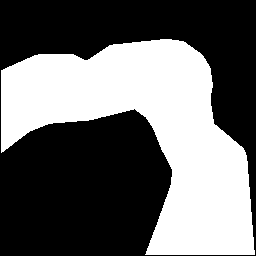} &
    		\includegraphics[width=0.22\textwidth]{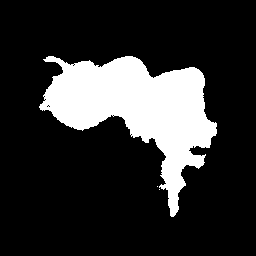} & 
    		\includegraphics[width=0.22\textwidth]{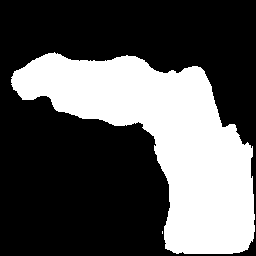} \\
    		(d) Ground Truth & (e) Output(DoubleUNet) & (f) Output(Ours) \\
    		\end{tabular}
    		}
    		\end{center}
    		\caption{Visual comparison of input images of NETWORK 2 in Triple-UNet and DoubleUNet. (a) is the input image and (d) is its corresponding ground truth. (b) and (c) are the inputs to NETWORK 2 in the corresponding algorithm. (e) and (f) are the prediction mask of DoubleUNet and the prediction mask of Triple-UNet, respectively.}
    		\label{Img3}
    	\end{figure*}

    	DoubleUNet only multiplies $\mathbf{x}^{(1)}$ and $\mathbf{u}$ to get Input2 $\mathbf{u}^{(2)}_{\mathrm{Multiply}}$, 
    	i.e. 
    	\begin{equation}
        \label{eq: multiply}
            \mathbf{u}^{(2)}_{\mathrm{Multiply}} =  \mathbf{x}^{(1)} \odot  \mathbf{u}.
        \end{equation}
    	Figure \ref{Img3} shows a visual comparison of the second input generated by DoubleUNet and our proposed network. From the visual point of view, the ROI partition of $\mathbf{u}^{(2)}_{\mathrm{ROIE}}$ is more obviously enhanced, than that of $\mathbf{u}^{(2)}_{\mathrm{Multiply}}$. In particular, the image $\mathbf{u}^{(2)}_{\mathrm{ROIE}}$ contains all information of the original image, and NETWORK 2 has the opportunity to correct the errors of NETWORK 1.
    	
    	In our proposed architecture, NETWORK 2 obtains a better score map $\mathbf{x}^{(2)}$ than $\mathbf{x}^{(1)}$ with features learned from NETWORK 1 and augmented input. However, NETWORK 2 may also misclassify some pixels belonging to the background as target objects. Hence we hope that NETWORK 3 correctly classifies these pixels, making the predicted score map of NETWORK 2 more refined and accurate. Therefore, we use the Multiply structure to get Input3 $\mathbf{u}^{(3)}_{\mathrm{Multiply}} = \mathbf{x}^{(2)} \odot \mathbf{u}$ as the input of NETWORK 3. $\mathbf{u}^{(3)}_{\mathrm{Multiply}}$ contains the target object area and the weakened background area. NETWORK 3 performs segmentation on $\mathbf{u}^{(3)}_{\mathrm{Multiply}}$ again to further fine-tune $\mathbf{x}^{(2)}$ for improving the segmentation accuracy, and then obtains the final score map $\mathbf{x}^{(3)}$.
	
\section{Experiment}
	To evaluate the effectiveness of the Triple-UNet architecture, we trained a variety of neural network architecture, including DoubleUNet, and UNet++. In this section, we present datasets, evaluation metrics, experimental setup and configuration, and data augmentation techniques used in all the experiments.
	
	\subsection{Datasets}
		We execute experiments on each model on the ISIC-2018 lesion boundary segmentation dataset \cite{2019Skin,2018The}. The dataset contains $2694$ dermatoscopy images of skin lesions and their ground truth. We use $80\%$, $10\%$, and $10\%$ of the data for training, validation, and testing, respectively. To improve GPU utilization and reduce training time, we adjust the size of each image to $256\times256$. In the training, we apply different data augmentation methods to the training set and validation set, including horizontal flip, vertical flip, Gaussian noise, blur, and random brightness contrast to increase the number of samples.
		
		The types of skin diseases in the lesion boundary segmentation dataset of ISIC-2018 mainly focus on melanoma, melanocytic nevus, and benign keratosis. 
		To verify the generalization ability of the model on other types of skin diseases, we test it on the ISIC-2019 dataset \cite{ISIC2019}. The training set of this dataset contains 25331 dermoscopic images with their diagnostic labels as ground truth, which include basal cell carcinoma (BCC), actinic keratosis (AK), dermatofibroma (DF), vascular lesion (VASC), and squamous cell carcinoma (SCC). 
		These diseases are not present in the ISIC-2018 dataset. 
		Unfortunately, ISIC-2019 is used for skin disease classification tasks, without pixel-level labels for segmentation, so we are unable to make a quantitative comparison. 
		Therefore, we only qualitatively analyze the image segmentation results on the training set of the ISIC-2019 dataset.
		
	\subsection{Implementation details}
    	All models are implemented and trained by using PyTorch $1.7.1$ framework on a single NVIDIA Tesla V100S GPU, 8-core CPU and 32GB RAM. We use binary cross-entropy as the loss function for all the networks, which is defined as:
    	\begin{equation}
    	  \label{lossfunction} L_{BCE}(\mathbf{x},\mathbf{y})=-\frac1N\sum_{i=1}^N(y_{i}\log{(x_{i})}+(1-y_{i})\log{(1-x_{i})}),
    	\end{equation}
    	where $\mathbf{x}$ is the predicted score map and its element $x_{i}$ represents the probability that pixel $i$ belongs to the estimated ROI partition, $\mathbf{y}$ is the corresponding ground truth and its element $y_{i}$ is the label ($1$ for positive points and $0$ for negative points) of pixel $i$ in the ground truth ROI partition.
    	
    	We use the Adam optimizer with its default parameter and weight decay of $0.0005$. The batch size is set to $16$ and the learning rate is set to $1e-5$. The learning rate decay applies an exponential learning rate schedule, $\gamma=0.98$. All models are trained for $100$ epochs. All models are trained from scratch except that VGG-19 in DoubleUNet is pre-trained on ImageNet.
		
	\subsection{Comparison with State-of-the-Arts}
    	\begin{figure*}[!t]
    	    \begin{center}
    	    \addtolength{\tabcolsep}{-5pt} 
    	    {%
    		\fontsize{8pt}{\baselineskip}\selectfont
    		\begin{tabular}{cccccc}
    		\includegraphics[width=0.16\textwidth]{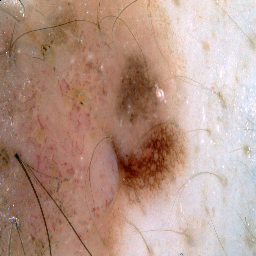} &
    		\includegraphics[width=0.16\textwidth]{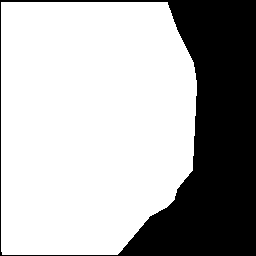} &
    		\includegraphics[width=0.16\textwidth]{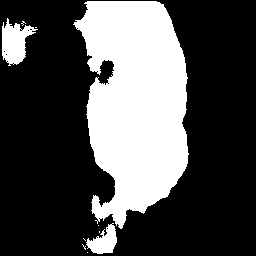} &
    		\includegraphics[width=0.16\textwidth]{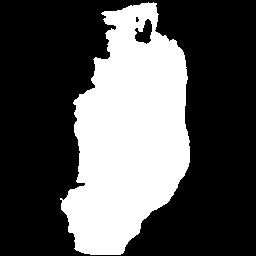} & 
    		\includegraphics[width=0.16\textwidth]{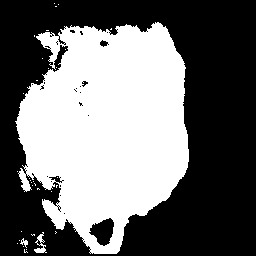} &
    		\includegraphics[width=0.16\textwidth]{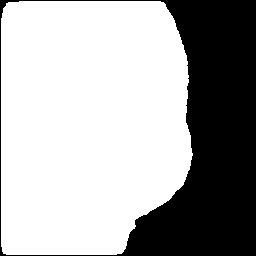} \\
    		
    		&  & Dice=0.619 & Dice=0.644 & Dice=0.747 & Dice=\textbf{0.975} \\
		    &  & mIoU=0.448 & mIoU=0.475 & mIoU=0.598 & mIoU=\textbf{0.952} \\
    		
    		\includegraphics[width=0.16\textwidth]{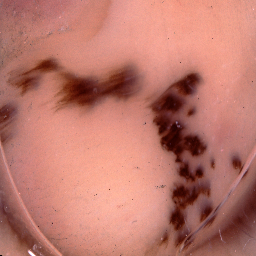} &
    		\includegraphics[width=0.16\textwidth]{res_DoubleUNet_30_label.png} &
    		\includegraphics[width=0.16\textwidth]{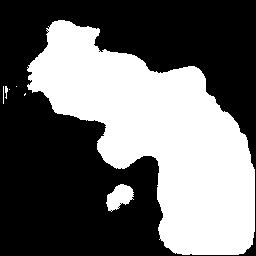} &
    		\includegraphics[width=0.16\textwidth]{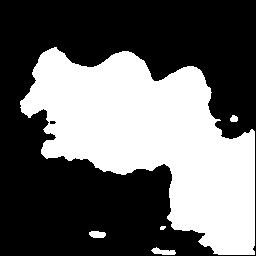} &
    		\includegraphics[width=0.16\textwidth]{res_DoubleUNet_30_output.png} &
         	\includegraphics[width=0.16\textwidth]{res_MultipleUNetP_30_output.png} \\
         	
         	&  & Dice=0.804 & Dice=0.785 & Dice=0.689 & Dice=\textbf{0.871} \\
		    &  & mIoU=0.674 & mIoU=0.648 & mIoU=0.540 & mIoU=\textbf{0.774} \\
         	
         	\includegraphics[width=0.16\textwidth]{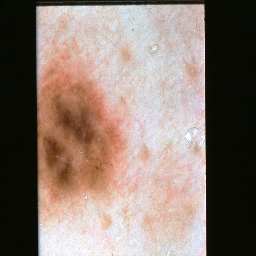} &
    		\includegraphics[width=0.16\textwidth]{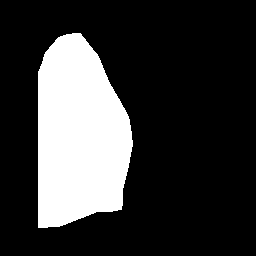} &
    		\includegraphics[width=0.16\textwidth]{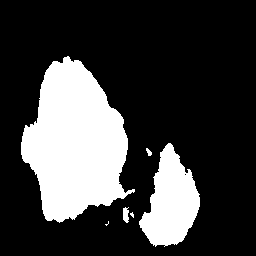} &
    		\includegraphics[width=0.16\textwidth]{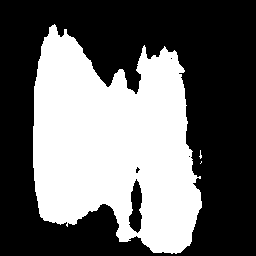} &
    		\includegraphics[width=0.16\textwidth]{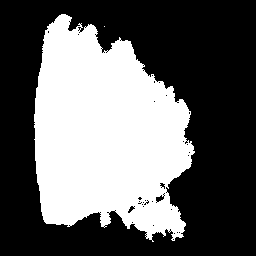} &
         	\includegraphics[width=0.16\textwidth]{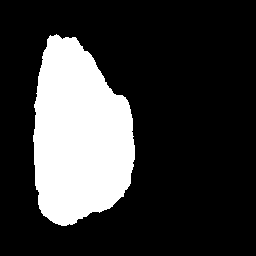} \\
         	
         	&  & Dice=0.834 & Dice=0.756 & Dice=0.792 & Dice=\textbf{0.964} \\
		    &  & mIoU=0.726 & mIoU=0.617 & mIoU=0.663 & mIoU=\textbf{0.931} \\
         	
         	\includegraphics[width=0.16\textwidth]{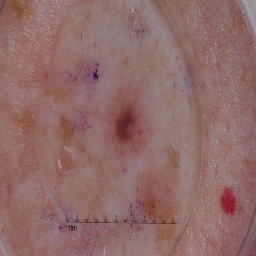} &
    		\includegraphics[width=0.16\textwidth]{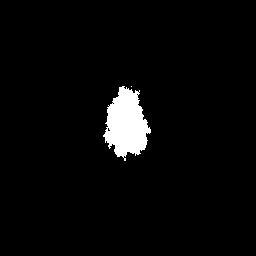} &
    		\includegraphics[width=0.16\textwidth]{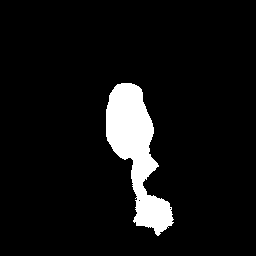} &
    		\includegraphics[width=0.16\textwidth]{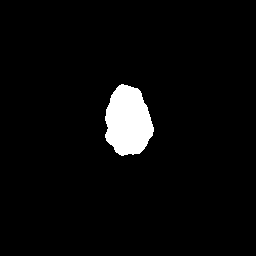} &
    		\includegraphics[width=0.16\textwidth]{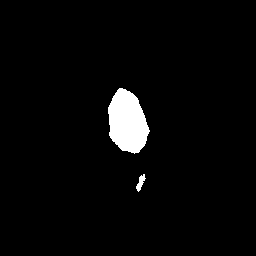} &
         	\includegraphics[width=0.16\textwidth]{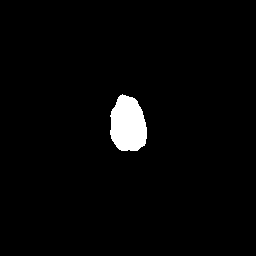} \\
         	
         	&  & Dice=0.801 & Dice=0.942 & Dice=\textbf{0.945} & Dice=0.922 \\
		    &  & mIoU=0.706 & mIoU=0.896 & mIoU=\textbf{0.900} & mIoU=0.864 \\
         	
         	\includegraphics[width=0.16\textwidth]{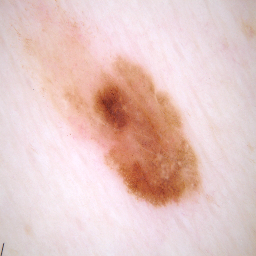} &
    		\includegraphics[width=0.16\textwidth]{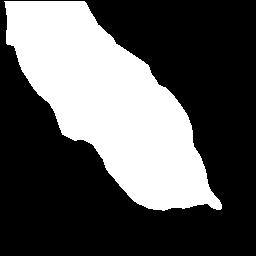} &
    		\includegraphics[width=0.16\textwidth]{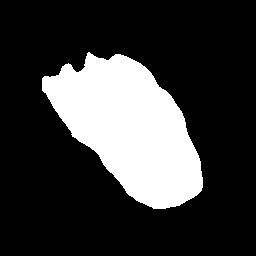} &
    		\includegraphics[width=0.16\textwidth]{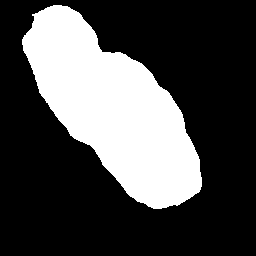} &
    		\includegraphics[width=0.16\textwidth]{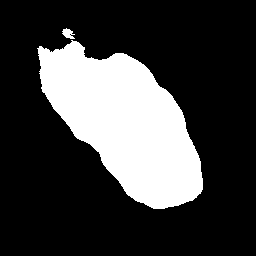} &
         	\includegraphics[width=0.16\textwidth]{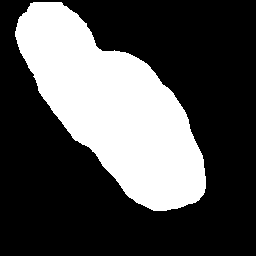} \\
         	
         	&  & Dice=0.849 & Dice=0.931 & Dice=0.877 & Dice=\textbf{0.950} \\
		    &  & mIoU=0.744 & mIoU=0.873 & mIoU=0.784 & mIoU=\textbf{0.905} \\
         	(a) Image & (b) Ground truth & (c) UNet++ & (d) UNet 3+ & (e) DoubleUNet & (f) Ours \\
    		\end{tabular}
    		}
    	    \end{center}
    	    \caption{Qualitative results on Lesion boundary segmentation dataset from ISIC-2018 \cite{2019Skin,2018The}}
    		\label{Img4}
    	\end{figure*}
	    
    	To compare the performance of Triple-UNet with state-of-the-arts on the lesion boundary segmentation dataset in more detail, the prediction results are usually classified into four categories: true positives (TP), false positives (FP), true negatives (TN), and false negatives (FN). We evaluated the standard measures of medical image segmentation such as mean Intersection over Union (mIoU), dice similarity coefficient (Dice), and Accuracy of all models. These evaluation metrics are defined as:
    	\begin{align}
    		\mathrm{mIoU} &=\frac{1}{2}\left(\frac{\mathrm{TP}}{\mathrm{TP}+\mathrm{FP}+\mathrm{FN}} + \frac{\mathrm{TN}}{\mathrm{TN}+\mathrm{FN}+\mathrm{FP}}\right),\\
    		\mathrm{Dice} &= \frac{2\mathrm{TP}}{2\mathrm{TP}+\mathrm{FP}+\mathrm{FN}},\\
    		\mathrm{Accuracy} &= \frac{\mathrm{TP}+\mathrm{TN}}{\mathrm{TP}+\mathrm{TN}+\mathrm{FP}+\mathrm{FN}.}
    	\end{align}
    	At the same time, the parameter quantity (param.), calculation speed (FLOPs), and reasoning speed (FPS) of the model are also the focus of our attention.
		
    	\begin{table}[t]
    		\centering
    		\caption{Quantitative results on Lesion boundary segmentation dataset from ISIC-2018 \cite{2019Skin,2018The}}
    		\label{Table 1}
    		\setlength{\tabcolsep}{3.5mm}{
    			\begin{tabular}{lcccccc}
    				\hline
    				Method & Dice & mIoU & Accuracy & param. & FLOPs & FPS \\
    				\hline
    				UNet++ \cite{2018UNet++} & 0.908 & 0.837 & 0.943 & 35M & 34.60G & 40/s \\
    				MultiResUNet \cite{2020MultiResUNet} & 0.909 & 0.839 & 0.945 & \textbf{27M} & \textbf{15.04G} & 39/s \\
    				AttentionUNet \cite{2018AttentionUNet} & 0.910 & 0.840 & 0.945 & 121M & 48.35G & \textbf{65/s} \\
    				DCUNet \cite{2020DCUNet} & 0.912 & 0.844 & 0.946 & 38M & 18.96G & 26/s \\
    				UNet 3+ \cite{2020UNet+++} & 0.914 & 0.846 & 0.946 & 103M & 197.97G & 13/s \\
    				DoubleUNet \cite{2020DoubleUNet} & 0.919 & 0.854 & 0.950 & 76M & 51.45G & 19/s \\
    				Triple-UNet & \textbf{0.925} & \textbf{0.865} & \textbf{0.953} & 87M & 40.22G & 13/s \\
    				\hline
    		\end{tabular}}
    	\end{table}
		
	    Table \ref{Table 1} shows the quantitative comparative results of our proposed Triple-UNet and other methods on the lesion boundary segmentation datasets. These methods include UNet++ \cite{2018UNet++}, AttentionUNet \cite{2018AttentionUNet}, DCUNet \cite{2020DCUNet}, UNet 3+ \cite{2020UNet+++} and DoubleUNet \cite{2020DoubleUNet}. Obviously, our model achieves the best segmentation performance. Our Triple-UNet achieves $0.925$ and $0.865$ in Dice and mIoU, respectively, which are $0.006$ and $0.011$ higher than DoubleUNet. Even compared with other state-of-the-art segmentation models, our Triple-UNet still achieves consistent and significant improvements on both metrics. In terms of scale and efficiency, although Triple-UNet is slightly more than DoubleUNet in terms of parameters, it is computationally superior to DoubleUNet because the number of filters in each layer of the encoder of Triple-UNet is half that of DoubleUNet. The visual quality is illustrated in Figure \ref{Img4}. 
	    One can see that for difficult and complex images, our proposed method segments the lesion fairly accurately while other algorithms fail to failed to identify the target objects. 
	    Taking the first image, for example, our segmentation mask is almost the same as the ground truth one, but the UNet++, UNet3+, and DoubleUNet segment is only half of the lesion area. 
	    However, for simple images with small objects, the results of our method are slightly lower than the other algorithms in terms of metrics, and the target area are slightly smaller than the other algorithms visually.
	    In summary, the qualitative and quantitative results confirm that our proposed network is more robust and has better performance than the state-of-the-arts.
	    
\begin{figure*}[!t]
    	    \begin{center}
    	    \addtolength{\tabcolsep}{-5pt} 
    	    {%
    		\fontsize{8pt}{\baselineskip}\selectfont
    		\begin{tabular}{m{1cm}<{\centering}m{2.5cm}<{\centering}m{2.5cm}<{\centering}m{2.5cm}<{\centering}m{2.5cm}<{\centering}m{2.5cm}<{\centering}}
    		BCC &
    		\includegraphics[width=0.16\textwidth]{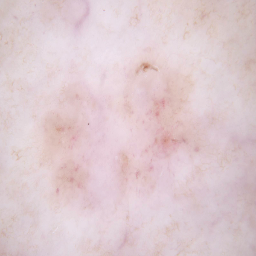} &
    		\includegraphics[width=0.16\textwidth]{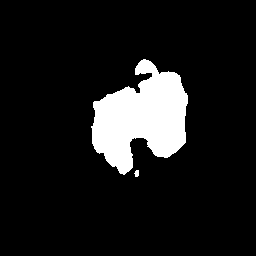} &
    		\includegraphics[width=0.16\textwidth]{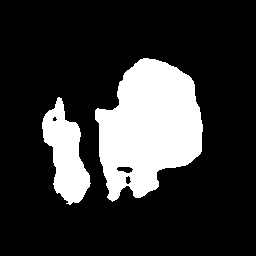} & 
    		\includegraphics[width=0.16\textwidth]{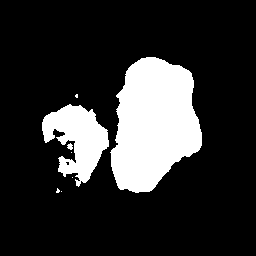} &
    		\includegraphics[width=0.16\textwidth]{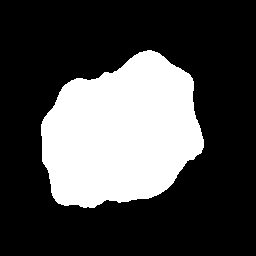} \\
    		
    		AK &
    		\includegraphics[width=0.16\textwidth]{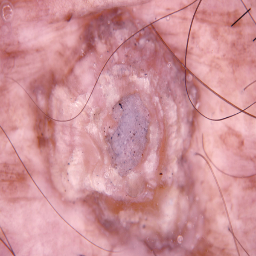} &
    		\includegraphics[width=0.16\textwidth]{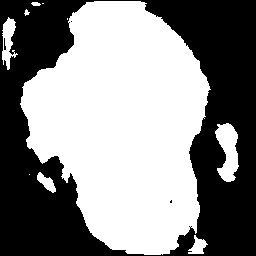} &
    		\includegraphics[width=0.16\textwidth]{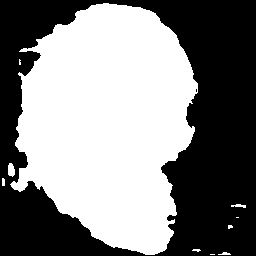} &
    		\includegraphics[width=0.16\textwidth]{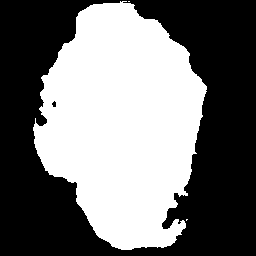} &
         	\includegraphics[width=0.16\textwidth]{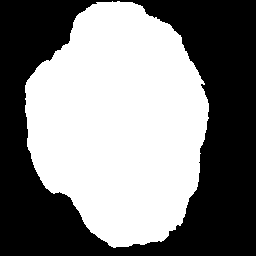} \\
         	
         	DF &
         	\includegraphics[width=0.16\textwidth]{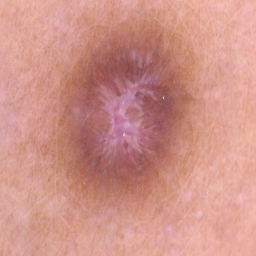} &
    		\includegraphics[width=0.16\textwidth]{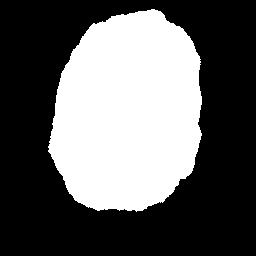} &
    		\includegraphics[width=0.16\textwidth]{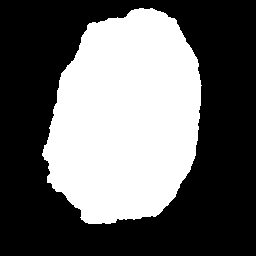} &
    		\includegraphics[width=0.16\textwidth]{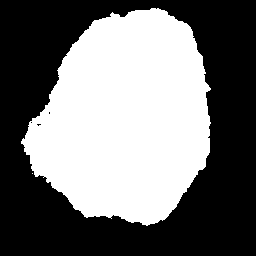} &
         	\includegraphics[width=0.16\textwidth]{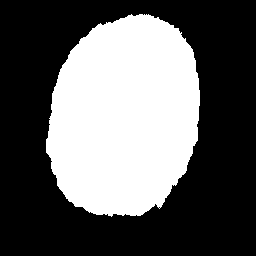} \\
         	
         	VASC &
         	\includegraphics[width=0.16\textwidth]{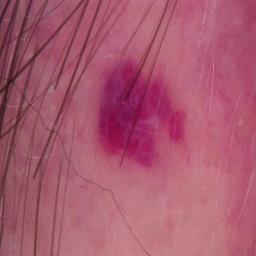} &
    		\includegraphics[width=0.16\textwidth]{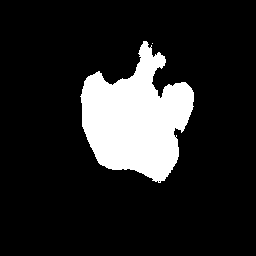} &
    		\includegraphics[width=0.16\textwidth]{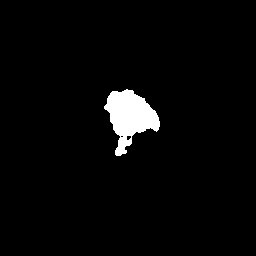} &
    		\includegraphics[width=0.16\textwidth]{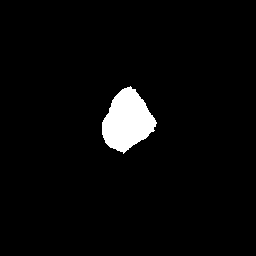} &
         	\includegraphics[width=0.16\textwidth]{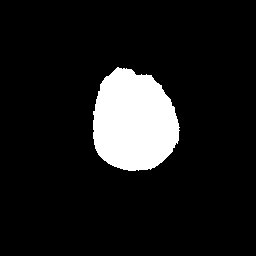} \\
         	
         	SCC &
         	\includegraphics[width=0.16\textwidth]{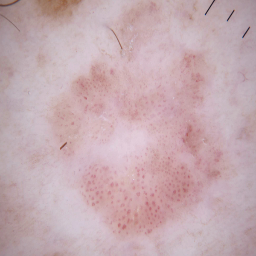} &
    		\includegraphics[width=0.16\textwidth]{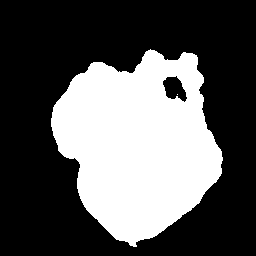} &
    		\includegraphics[width=0.16\textwidth]{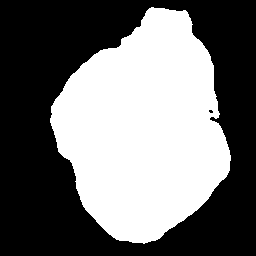} &
    		\includegraphics[width=0.16\textwidth]{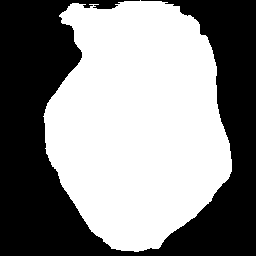} &
         	\includegraphics[width=0.16\textwidth]{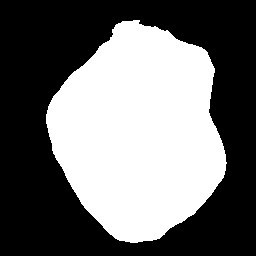} \\
         	 & (a) Image & (b) UNet++ & (c) UNet 3+ & (d) DoubleUNet & (e) Ours \\
    		\end{tabular}
    		}
    	    \end{center}
    	    \caption{Segmentation results on ISIC-2019 \cite{ISIC2019}. BCC is basal cell carcinoma. AK is actinic keratosis. DF is dermatofibroma. VASC is vascular lesion. SCC is squamous cell carcinoma.}
    		\label{Img5}
    	\end{figure*}

	    The segmentation results of the model on the ISIC-2019 dataset are shown in Figure \ref{Img5}. 
	    Although precise labeling by physicians was not obtained, it is observed visually that the segmentation results of our method have less noise, smoother edges and more intuitive segmentation.
        Our method has better generalization ability for previously unseen skin disease types.
	    
    \subsection{Ablation Study}
        We conduct ablation studies to demonstrate the effectiveness of two major components in Triple-UNet: three U-Shaped network integration and ROIE. Table \ref{Table 2} summarizes the experimental results for different arrangements. DoubleUNet contains two U-shaped sub-networks connected by a Multiply structure, and Triple-UNet contains three U-shaped sub-networks connected by ROIE and Multiply in turn. DoubleUNet*, Triple-UNet-a, Triple-UNet-b, and Triple-UNet-c are segmentation models generated by removing a sub-network or changing the connection structure between sub-networks or changing the order of the connection structure between sub-networks on the basis of Triple-UNet. 4-UNet and 5-UNet are integrated with four and five U-Shaped networks similar to Triple UNet. DoubleUNet* applies the ROIE module, which improves the performance of the network even though we do not use the ASPP module.
        This result shows that our proposed ROIE module helps the network achieve better segmentation performance compared to the Multiply structure. Triple-UNet-a and Triple-UNet-b increase the number of integrated U-shaped sub-networks to $3$ and only use the Multiply structure and ROIE structure, respectively, and they both have substantial improvements in Dice and mIoU. Integrating multiple U-shaped networks improves the performance of the model. However, simply increasing the number of inherited networks does not help to improve performance, such as 4-UNet and 5-UNet. Perhaps other special designs are needed. In addition, considering resource consumption, we do not continue to increase the number of sub-networks. Contrary to the order in which Triple-UNet is used, Triple-UNet-c uses the Multiply structure first and then uses the ROIE module, and its performance is much worse than that of Triple-UNet. From the results, we find that using the ROIE module and the Multiply structure in turn, enhancing the input first and then performing the refined segmentation, is the best arrangement strategy. 
        
    	\begin{table*}[t]
    		\centering
    		\caption{Result on Lesion boundary segmentation dataset from ISIC-2018 \cite{2019Skin,2018The}}
    		\label{Table 2}
    		\setlength{\tabcolsep}{4pt}{
    			\begin{tabular}{llcccccccc}
    				\hline
    				Method & Connection Structure & Dice & mIoU & Accuracy & param. & FLOPs & FPS \\
    				\hline
    				DoubleUNet \cite{2020DoubleUNet}  & Multiply & 0.919 & 0.854 & 0.950 & \textbf{76M} & \textbf{51.45G} & \textbf{19/s} \\
    				DoubleUNet* & ROIE & 0.920 & 0.856 & 0.950 & 76M & 51.45G & 19/s \\
    				Triple-UNet-a & Multiply + Multiply & 0.922 & 0.860 & 0.951 & 87M & 40.22G & 13/s \\
    				Triple-UNet-b & ROIE + ROIE & 0.922 & 0.859 & 0.951 & 87M & 40.22G & 13/s \\
    				Triple-UNet-c & Multiply + ROIE & 0.915 & 0.848 & 0.947 & 87M & 40.22G & 13/s \\
    				4-UNet & 2*Multiply + ROIE & 0.922 & 0.859 & 0.951 & 118M & 54.45G & 9/s \\
    				5-UNet & 3*Multiply + ROIE & 0.917 & 0.852 & 0.948 & 148M & 68.67G & 7/s \\
    				Triple-UNet & ROIE + Multiply & \textbf{0.925} & \textbf{0.865} & \textbf{0.953} & 87M & 40.22G & 13/s \\
    				\hline
    		\end{tabular}}
    	\end{table*}
	
	\section{Conclusion}
In this paper, we propose a new CNN-based architecture called Triple-UNet, which is used to accurately segment skin lesions from dermoscopy images. The model uses the combination of three UNet architectures and takes advantage of multiple segmentation and attention modules. Triple-UNet extracts more efficient features from the input image and adjusts the input image to a form more suitable for the network in segmentation. We evaluated our model on the ISIC-2018 dataset. The results show that Triple-UNet outperforms other SOTA models in Dice and mIoU. In the future, we will focus on optimizing the architecture of Triple-UNet, designing a simplified structure with fewer parameters to accelerate the segmentation of the model while maintaining its capabilities and making it suitable for more medical image segmentation tasks. 
	
	\vskip 10mm
	
	\noindent $^1$ Inner Mongolia University, Hohhot 010020, China.\\
	\indent  Email: lqgoqk@aliyun.com, yuguomath@aliyun.com, wucaiyingyun@163.com, cgq@imu.edu.cn, qyjin2015@aliyun.com
	
	\noindent $^2$ Inner Mongolia Normal University, Hohhot 010020, China.\\
    \indent  Email: saheya@imnu.edu.cn
	
\end{document}